\documentstyle[epsf,12pt]{article}

\newcommand{\mb}[1]{ { \mbox{\boldmath{$#1$}}}  }

\begin{document}

\begin{center} 
{\Large Effect of Disorder on a D--wave Superconductor} \\~ \\
Grzegorz Litak \\~
\\ Department of Mechanics, Technical University of Lublin,
Nabystrzycka 36,
20-618 Lublin, Poland 

\end{center}


\begin{abstract} 
We apply the Coherent Potential Approximation (CPA) to
an extended Hubbard model to describe
 disordered superconductors with
$d$--wave
pairing.
We discuss the 
pair--breaking effect 
caused by
non--magnetic
disorder
in presence 
of Van Hove singularity. 
\end{abstract}

\section{Introduction} 
The original treatments of the
influence
of magnetic and non--magnetic disorder on superconductors
applied to   
classic BCS
superconductors \cite{And59}, has been reexamined for
superconductors 
 with $d$--wave symmetry of order parameter
\cite{Gor83,Gho01,Lit98,Har00}.
On the other hand, in high temperature superconductors 
the distance between the chemical potential and Van Hove singularity
was found to be relatively small.   
This has lead to formulation of
Van Hove scenario for high temperature superconductors   
which
says that optimal critical temperature is reached when chemical potential
passes through the Van Hove singularity in the
density of states \cite{Mar97}.
However, doping with charge carriers does not only
change the density
in the system but also smear the density of states
eliminating its singularities
and,  specially
for anisotropic superconductors, introduce the electron pair--breaking
phenomenon \cite{Gor83,Har00,Lit98}.

\section{Superconductivity in a Disordered System}
We start with the single band Hubbard model with an attractive
extended
interaction which is described by the following Hamiltonian
\cite{Mic88}:

\begin{equation}
\label{eq1}
H=\sum_{ij \sigma} t_{ij}c^{\dagger}_{i \sigma} c_{j \sigma} +
\frac{1}{2} \sum_{ij} U_{ij} n_i n_j - \sum_{i}(\mu-\varepsilon_i)n_i.
\end{equation}
In the above $n_i=n_{i \uparrow}+n_{i \downarrow}$ is the charge on  site
labeled $i$, $\mu$ is the
chemical
potential. Disorder is introduced into the problem by allowing the local
site
energy $\varepsilon_i$ to vary randomly from site to site, $c^{\dagger}_{i
\sigma}$ and $c_{i \sigma}$ are the Fermion
creation and annihilation operators for an electron
 on site $i$ with spin $\sigma$,
$t_{ij}$ is the amplitude for      
hopping from site $j$ to site $i$
and finally  $U_{ij}$ is the attractive interaction ($U_{ij} < 
0$),  between electrons on neighbour sites ($i \ne j$).

\begin{figure}[htb]
\leavevmode
\vspace*{-4.5cm}
\hspace*{0.5cm}
\epsfxsize=4.0cm
\epsffile{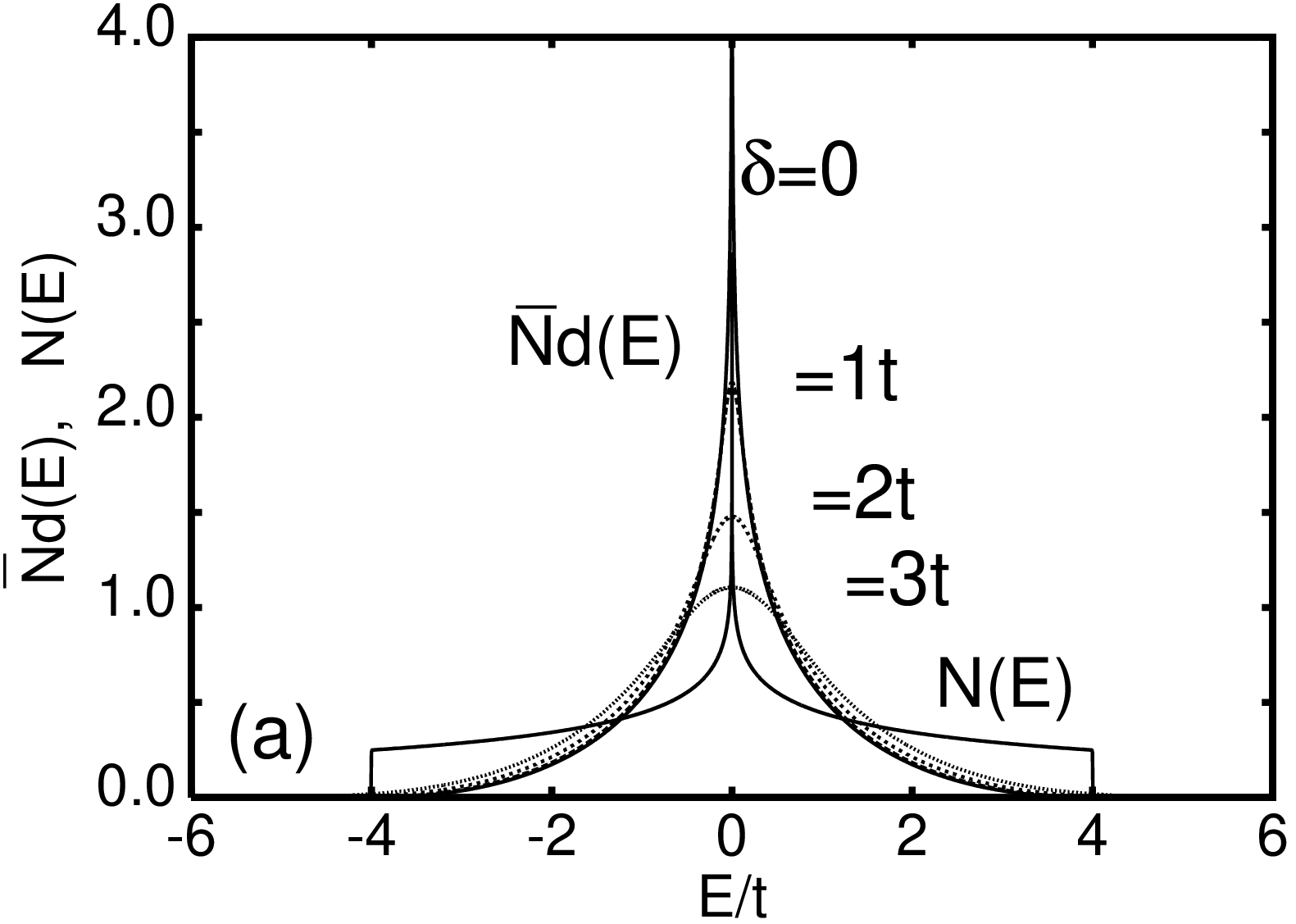}
\epsfxsize=4.0cm
\hspace*{2.5cm}
\epsffile{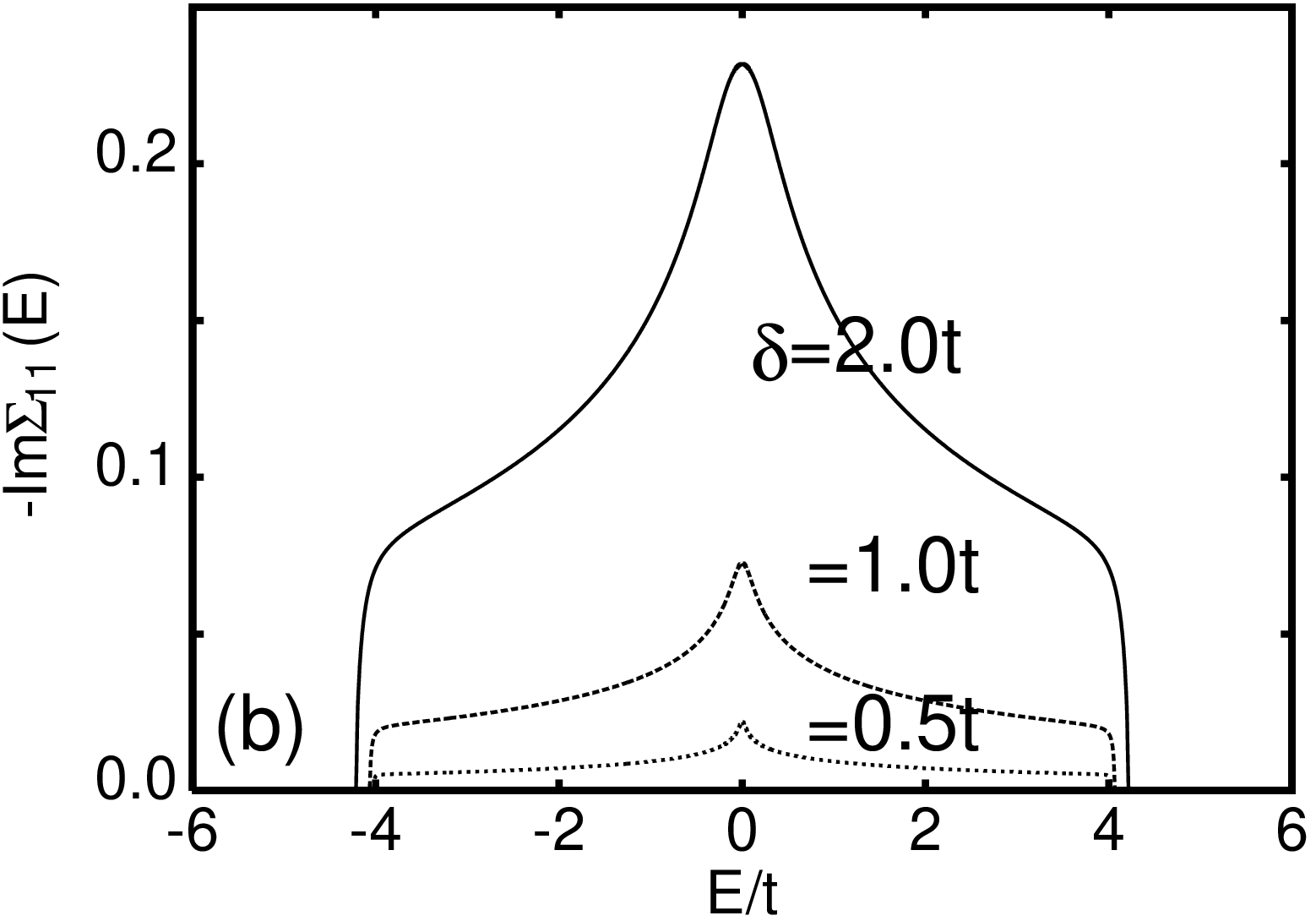}
\vspace{2.8cm}

\caption[Fig. 1]{ ($a$) 
Projected $d$--wave density of states $\overline{N_d}(E)$ in presence of
disorder for various $\delta$ and the electron density of states $N(E)$ 
for 
the clean system.
($b$) $-{\rm Im}\Sigma_{11}(E)$
for various band fillings $\delta$.
}
\end{figure}

Here we will assume, for simplicity, that the random site energy
$\varepsilon_i$ has a uniform distribution $\varepsilon_i \in
[-\delta/2,\delta/2]$.
Following the usual way we shall apply the Coherent
Potential Approximation (CPA) to mean that the coherent potential  $\mb
\Sigma(E)=\mb \Sigma(i,i;E)$ \cite{Lit98},
in a site
approximation, is defined by the zero value of an averaged  t-matrix $ \mb
T(i,i;
E)$.

The linearized gap equation in Hartree-Fock-Gorkov approximation
\cite{Lit98,Mic88}:
\begin{equation}
1= \frac{|U|}{\pi} \int^{\infty}_{- \infty} {\rm d} E~ {\rm tanh}
\frac{E}{2k_BT_c }~
{\rm Im} \frac{ \overline G^d_{11} (E)}{ 2 E - {\rm Tr} \mb \Sigma (E)},
\end{equation}
where $T_c$ is the critical temperature, $k_B$ denotes the Boltzmann
constant, $\overline G^d_{11} (E)$ is an
averaged electron Green function
 which defines the
weighted density of states (DOS) of
$d$-wave electron states $\overline N_d( E)$:
\begin{equation}
\overline N_d(E)= - \frac{1}{\pi} {\rm Im} \overline G^d_{11} (E) = -
\frac{1}{\pi
N}
\sum_{ \small \mb k}
{\rm Im} \frac{\eta_{\mb k}^2}{4}\frac{1}{E - \Sigma_{11}(E) -
\varepsilon_{\small \mb k} + \mu}.
\end{equation}
$\Sigma_{11}(E)$ describes the electron
self energy in the  normal disordered system and $\eta_{\mb k} =
2(\cos{ k_x} -cos{k_y})$.
Examples of projected densities of states $N_d(E)$ for disordered 2D
system stem are presented in Fig. 1a while the corresponding self energies
are plotted in Fig. 1b.  
\begin{figure}[htb]
\leavevmode
\vspace*{-4.5cm}
\hspace*{0.5cm}
\epsfxsize=4.0cm
\epsffile{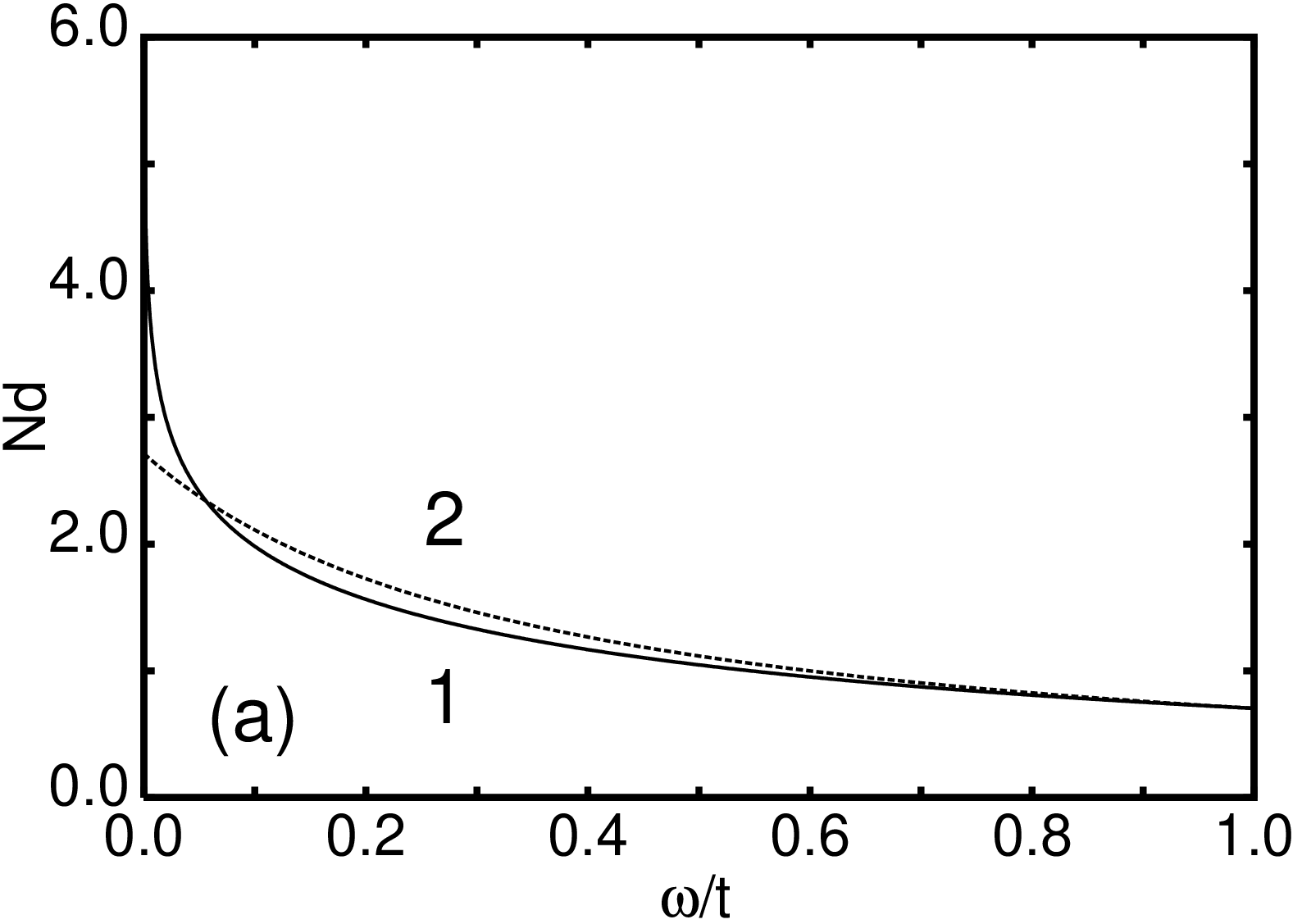}
\epsfxsize=4.0cm
\hspace*{2.5cm}
\epsffile{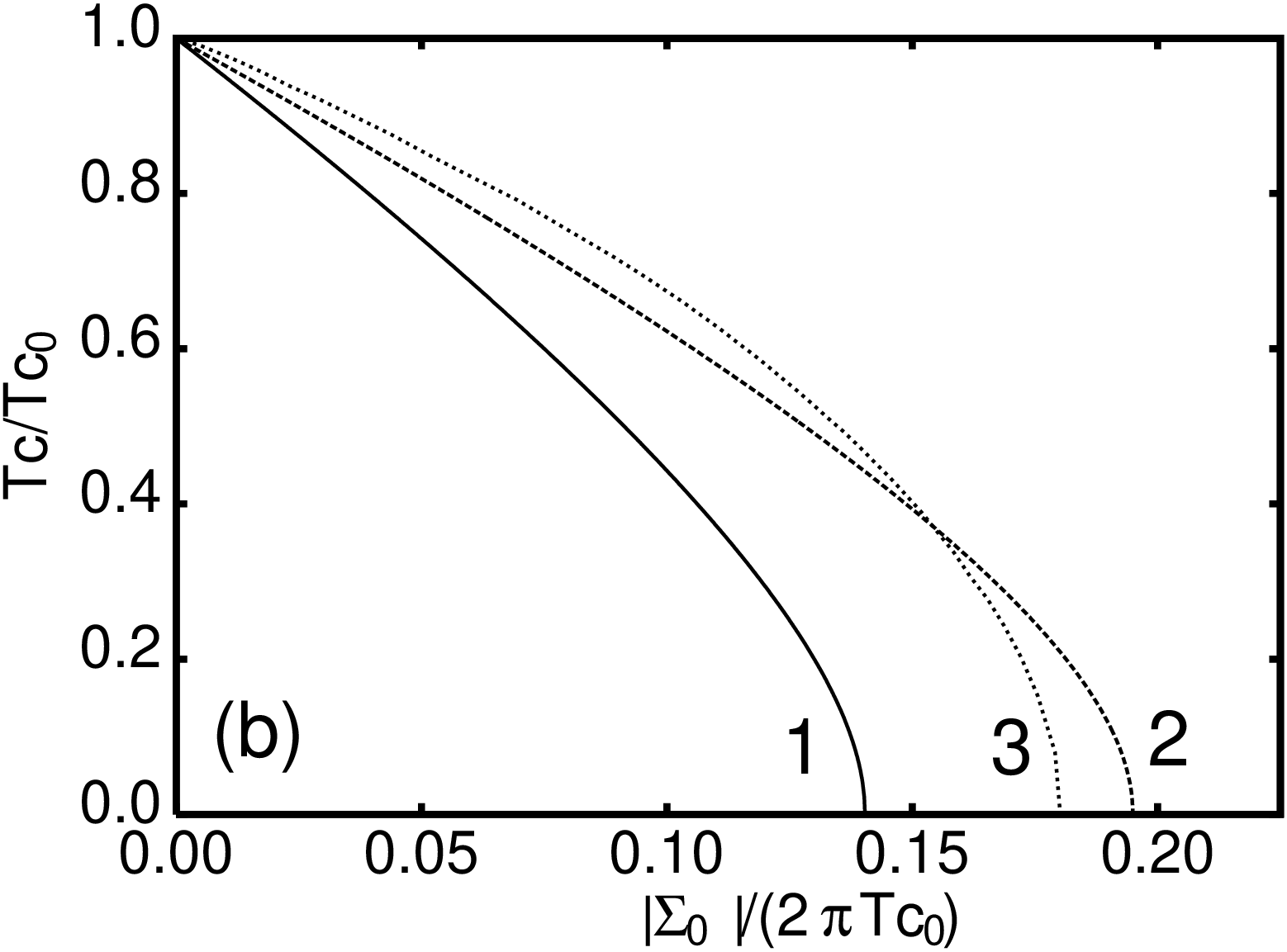}
\vspace{2.8cm}

\caption[Fig. 2]{ ($a$)
$N_d(\imath \omega)$ versus $\omega$ - full line '1', the fitting curve
$c/(\omega+b)$ for
$c=0.95$ and $b=0.35$
- dashed line '2'.
($b$) $T_c$ versus $|\Sigma_0|$: '1' the standard Abrikosov--Gorkov
formula, '2' results obtained from Eq. 5, '3' results 
from numerical solving of the gap equation (Eq. 2) for $U=-2t$.    
}
\end{figure}
Equation 1 can be rewritten in terms of 
Matsubara frequencies $\omega_n$ as

\begin{equation}
1= \frac{|U|}{2 k_B T_c} \sum_n \frac{ \overline G^d_{11}
(\imath \omega_n)}{ 2 \imath \omega_n 
- {\rm Tr} \mb \Sigma (\imath \omega_n)}
\end{equation}

In Fig. 2a
we show the density $N^d(\imath \omega)=-{\rm Im} \overline G^d(\imath
\omega_n)/\pi 
$ 
 versus
imaginary frequency $\omega$,
where $E=\epsilon_f+\imath
\omega$ for  Fermi energy $\epsilon_f$ chosen at Van Hove singularity
($\epsilon_f=0$, Fig. 1a).
In the standard treatments [4] it was assumed to be constant. However in
our case, due to the Van Hove  singularity, it depends strongly on
$\omega$.
The corresponding
projected density 
$N^d(\imath \omega)$ can be roughly approximate by 
a simple formula
$N_d(\imath \omega)= c/ (\omega + b)$
where $c$ and  $b$ are constants (Fig. 2a). 
Approximating also $\Sigma(i \omega)= -\imath |\Sigma_0| {\rm sgn}
(\omega)$ 
we get from Eq. 2 the analytic pair-breaking formula: 
\begin{equation}
\psi \left(\frac{1}{2}\right) -\psi \left( \frac{1}{2} +
\frac{b}{2 \pi
T_{c0}}\right)=
\psi \left( \frac{1}{2} +  \rho_c\right) -
\psi \left( \frac{1}{2} + \rho_c + \frac{b}{2 \pi
T_{c}}\right),
\end{equation}
where 
$\rho_c = |\Sigma_0|/(2 \pi T_{c})$ is a pair--breaking parameter
and $T_{c0}$ is the critical temperature for a clean superconductor.
Note that for large $b$ ($b \rightarrow \infty$) Eq. 4 transforms into
the standard
Abrikosov-Gorkov equation \cite{And59}. Fig. 2b presents the comparison
between these analytic formulae as well as  the numerical results obtained
from the 
gap equation (Eq. 2). Both  analytic and numerical results (Fig. 2b) show
that
Van Hove singularity make the superconducting system more robust in
presence the disorder.

\section{Conclusions}

Analyzing  the effect of disorder on disordered $d$--wave superconductor
we have found the additional influence of the  Van
Hove
singularity.
In the presence of a weak disorder $\delta < t$ we observe only small
change
of the projected density of
states $\overline N_d$ (Fig. 1a) present in the gap equation (Eqs. 2,4)
In the same time we observe  the rapid
degradation of $T_c$, which is
connected 
with the pair-breaking effect (Eq. 2).
Interestingly, Van Hove singularity modifies the 
standard Abrikosov-Gorkov formula, originally obtained for a constant
density
of states, increasing the critical value of $|\Sigma_0|$
which destroys the superconducting phase (Fig. 2b).
That result is in qualitative agreement with the experimental results
on Zn substitutions \cite{Kar00}.

\section*{Acknowledgements}
 The author would
like to thank  
Prof. K.I. Wysokinski, Prof. B. L. Gy\"{o}rffy 
and Dr. J.F. Annett  for
helpful discussions.

\end{document}